\documentclass[aps,prb,twocolumn,groupedaddress]{revtex4}

\newcommand{\centigrade}{{$^\circ$\textrm{C}}}
\usepackage{graphicx}

\begin{document}

\title{Electron conduction through quasi-one-dimensional indium atomic wires on silicon}

\author{Takashi Uchihashi$^{\mathrm{a)}}$}

\author{Urs Ramsperger$^{\mathrm{b)}}$}

\affiliation{National Institute for Materials Science, 1-2-1, Sengen, Tsukuba, Ibaraki 305-0047, Japan}

\date{\today}

\begin{abstract}
Electron conduction through quasi-one-dimensional (1D) indium atomic wires on silicon (the Si(111)-4$\times 1$-In reconstruction) is clarified with the help of local structural analysis using scanning tunneling microscopy. 
The reconstruction has a conductance per square as high as 100 $\mu$S, with global conduction despite numerous surface steps. 
A complete growth of indium wires up to both the surface steps and the lithographically printed electrodes is essential for the macroscopic transport.
The system exhibits a metal-insulator transition at 130 K, consistent with a recent ultraviolet photoemission study [H. W. Yeom, S. Takeda, E. Rotenberg, I. Matsuda, K. Horikoshi, J. Schaefer, C. M. Lee, S. D. Kevan, T. Ohta, T. Nagao, and S. Hasegawa, Phys. Rev. Lett. \textbf{82}, 4898 (1999)].

\end{abstract}.

\maketitle 
The downsizing of microelectronics in pursuit of high density and high speed has brought them into the nanoscale regime in terms of their dimension.
Suppose that the components of nanoelectronics, fabricated on a substrate, are further reduced in size to an extent that they are monoatomically thin.
These are nothing else than surface atomic structures, which often lead to reconstructions whose crystallographic and electronic properties are essentially distinct from their original bulk materials.\cite{Lifshits_SurfacePhases}
Thus, understanding electron conduction properties of surface structures is of fundamental importance in the future nanoelectronics.
In spite of its long history,\cite{Henzler_SurfaceConduction,Henzler_AgSi,Bauer_Pbfilm} interest in electron transport phenomena originating from well-defined surface reconstructions has been renewed only recently.\cite{Hasegawa_Agroot3,Weitering_Si100}
To measure such conduction, two main difficulties must be overcome: 1) separating out conduction through the underlying subsurface space charge layer \cite{Henzler_SurfaceConduction} and 2) establishment of reliable electrical contacts to the atomic structures that are stable over a wide temperature range. 
Furthermore, as electron conduction through surface atomic structure is assumed to be grossly affected by defects, local structural analysis is essential to clarify their unambiguous transport properties. 

In this paper an unprecedented combination of conductivity measurements and scanning tunneling microscopy clarifies electron transport properties of the Si(111)-4$\times 1$-In reconstruction. 
This system is particularly interesting because it consists of quasi-one-dimensional (1D) indium wires on a silicon surface.\cite{Nogami_In4x1,Abukawa_In4x1,Netzer_In4x1,Yeom_In4x1,Bunk_In4x1,Kumpf_In4x1}
We extract surface conduction through the comparison of two surface structures, one of which includes intentionally introduced defects. 
Indium atomic wire arrays have a global conductance per square as high as 100 $\mu$S despite numerous surface steps. 
A complete growth of atomic wires up to both the surface steps and the lithographically printed electrodes is found to be essential for conduction over macroscopic lengths.
We find a metal-insulator transition at 130 K, consistent with a recent ultraviolet photoemission study.\cite{Yeom_In4x1}

\begin{figure}[tbp]
\begin{center}\leavevmode
\includegraphics[width=0.86\linewidth]{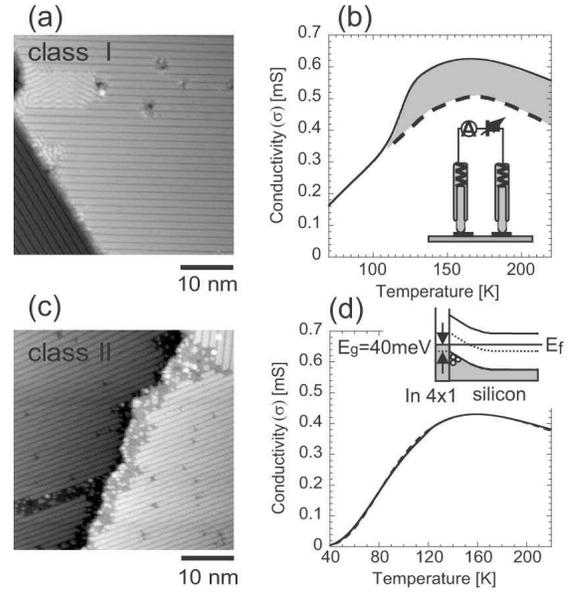}
\caption{
(a) A representative STM image of the class I samples. 
(b) A representative temperature dependence of conductivity ($\sigma$) of the class I samples (solid line).
The dashed line is the speculated temperature dependence of $\sigma$ above 110 K due to the contribution of the subsurface space charge layer. 
Inset: a diagram of the electrical conduction measurement of the Si(111)-4$\times 1$-In reconstruction.
(c) A representative STM image of the class II samples. 
(d) A representative temperature dependence of conductivity ($\sigma$) of the class II samples (solid line).  
The variation of $\sigma$ can be well described by the equation $\sigma = C \exp(-E_{g}/k_{B}T)T^{-\alpha}$ (dashed line). 
Inset: the energy band diagram of the system of the surface Si(111)-4$\times 1$-In reconstruction and the silicon substrate.
}
\label{fig1}
\end{center}
\end{figure}

All experiments are performed under ultrahigh vacuum (UHV) conditions. Non-doped Si(111) (resistivity $\rho > 1000 \ \Omega \ \mathrm{cm}$) is chosen as the substrate to minimize the electron conduction through the bulk. 
Tantalum, a typical refractory metal, is adopted as the electrode material to avoid diffusion during high temperature sample treatments. 
Two electrode pads separated by 1 mm are deposited beforehand on the sample with an electron beam evaporator (the inset of Fig. 1(b)). After loading into the UHV chamber, the samples are cleaned by flashing at 1150 \centigrade\ for 30 s. Indium is then deposited to a thickness of 1.8 monolayers (ML), followed by annealing around 450 \centigrade\ for 5 minutes. 
This develops indium atomic wire arrays on a silicon surface (Si(111)-4$\times 1$-In reconstruction).\cite{Nogami_In4x1,Abukawa_In4x1,Netzer_In4x1,Yeom_In4x1,Bunk_In4x1,Kumpf_In4x1}
The 4$\times 1$ reconstruction is confirmed over an extensive area by Low Energy Electron Diffraction (LEED). 
Local structures are observed with STM at room temperature, revealing defects in the wires, surface steps and domain boundaries. 
We find two classes of samples in terms of structural growth; I) complete and II) incomplete growth of indium atomic wires near the surface steps. 
Figures 1(a) and 1(c) show representative STM images respectively.\cite{Another_Phase}
After indium deposition, annealing at a temperature slightly higher than that required for complete growth strips off indium from the step edges. 
Thus, growth of the reconstruction at steps is controllable, and is homogeneous within the area of electron conduction. 

Voltage-biased dc two-probe measurements are conducted after STM observation. 
Two spring-loaded gold-coated probes are pressed onto the electrodes to insure stable and reliable electrical contacts (the inset of Fig. 1(b)). 
The current-voltage ($I-V$) characteristics are linear for $-1 \ \mathrm{V} < V < 1 \ \mathrm{V}$ over a wide temperature range, confirming the absence of a Schottky barrier at the electrode interfaces. 
The temperature dependence of conductivity is measured between room temperature and 6 K.
No contamination is apparent in the STM images after the cooling cycles. 
The carriers in the bare silicon substrates are found to be quenched below 220 K. This allows the selective detection of electrical conduction originating from the surface structures: the possible conduction path is either the surface reconstruction or the underlying subsurface space charge layer.\cite{Henzler_SurfaceConduction,Hasegawa_Agroot3} 

We clarify electron conduction through the Si(111)-4$\times 1$-In reconstruction through the comparison of the two sample classes (I, II). 
Figure 1(b) shows the temperature dependence of conductivity ($\sigma$) of a representative class I sample (solid line).\cite{footnote1} 
$\sigma$ exhibits a broad maximum around 160 K with decreasing temperature, followed by a significant drop below 130 K. 
This drop indicates a dramatic change in the conduction mechanism. 
Complete growth of indium wires right up to the surface step (Fig. 1(a)) suggests that conduction is through both the surface reconstruction and the subsurface space charge layer in the class I samples. 
Destruction of the indium wires at steps edges should eliminate the surface conduction, leaving only the subsurface conduction. 
Figure 1(d) exemplifies the temperature dependence of conductivity of the class II samples. 
While the overall behavior is similar to that of the class I samples, the sudden decrease in $\sigma$ below 130 K is absent.
The value of $\sigma$ above 130 K is also significantly lower than that of the class I samples (see the dashed line in Fig. 1(b) as an eye guide).
This difference is not due to change in space charge layer conduction caused by different structural growth at steps, because this would lead to a rather uniform shift in $\sigma$ across the entire temperature region.
Thus we conclude that the gray region of Fig. 1(b) is the contribution from the surface atomic wires; the conductivity is approximately 100 $\mu$S. 
The sudden decrease in $\sigma$ below 130 K is ascribed to a change in the surface conduction mechanism.

\begin{figure}[tbp]
\begin{center}\leavevmode
\includegraphics[width=0.88\linewidth]{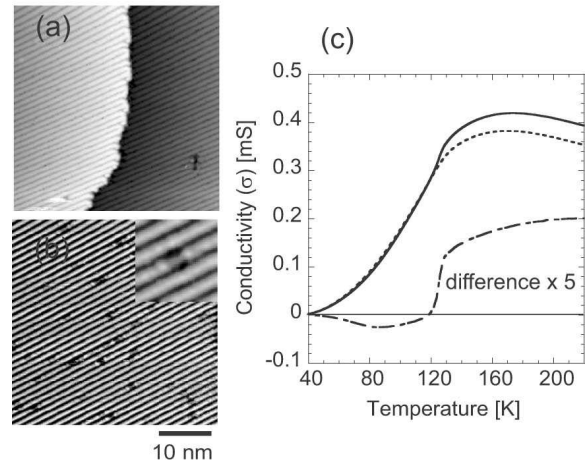}
\caption{
(a) An STM image of the sample with the Si(111)-4$\times 1$-In reconstruction  in the first measurement. 
(b) An STM image of the sample in the second measurement. Inset: magnified image of an indium-induced defect.
(c) The temperature dependence of conductivity of a sample with the Si(111)-4$\times 1$-In reconstruction in two consecutive measurements.
Solid line: first measurement on the sample with a complete reconstruction (see (a)). Dashed line: second measurement on the sample with defects in the middle of the wires (see (b)).  
The difference between the two measurements is the contribution from the surface reconstruction (dotted-dashed line, five-times magnified). 
}
\label{fig2}
\end{center}
\end{figure}

We confirm that the residual conductivity found in the class II samples (Fig. 1(d)) is the contribution of the subsurface space charge layer. 
The variation of $\sigma$ can be well described by the equation 
\begin{equation}
\sigma = C \exp(-E_{g}/k_{B}T)T^{-\alpha}                     
\end{equation}
where $C$ is a constant and $E_{g}$ and $\alpha$ are parameters. 
By curve fitting, one obtains $E_{g} = 39.7 \mathrm{meV}$  and $\alpha = 2.89$. 
The fitting curve (dashed line in Fig. 1(d)) fully supports the experiment.
The diagram of Fig. 1(d) (inset) illustrates the origin of these parameters.\cite{Henzler_SurfaceConduction} 
Because of the different Fermi levels between the surface Si(111)-4$\times 1$-In reconstruction and the silicon substrate, the silicon bands bend upwards near the surface. 
This induces hole carriers in the subsurface space charge layer. 
$E_{g}$ is attributed to the energy difference between the Fermi level and the upper bound of the silicon valence band near surface, representing the activation energy for holes. The term $T^{-\alpha}$ in equation (1) corresponds to the power-law $T$-dependence of the hole mobility. 
The value $E_{g} = 39.7 \mathrm{meV}$ is consistent with a previous report,\cite{Abukawa_In4x1} and the value $\alpha = 2.89$ is close to $\alpha = 2.20, 2.42$ for p- and n-type silicons respectively.\cite{Sze_Semiconductor} Once the band bending is determined, one can estimate the conductivity ($\sigma$) of the space charge layer using the reported silicon mobility.\cite{Hasegawa_Agroot3,Handbook_ChemPhys} 
At room temperature, $\sigma$ is estimated to be 170 $\mu$S, consistent with the measured value 300 $\mu$S considering that the conductivity of the substrate (70 $\mu$S) is included at room temperature. 
Thus we conclude that the electron conduction in the class II samples is dominated by the subsurface space charge layer. 

We have demonstrated electron conduction through surface reconstruction by intentionally destroying wires at steps, while their quantitative temperature dependence remains undetermined. 
To clarify this, we extract surface conduction by introducing defects in the middle of the wires on the \textit{same} sample. 
A metal-insulator transition of this system is observed via transport experiments. 
First, a complete Si(111)-4$\times 1$-In reconstruction is formed with negligible defects (Fig. 2(a)). 
The conductivity $\sigma$ shows a significant decrease at 130 K (Fig. 2(c), solid line). After returning to room temperature, a small amount of indium (0.04 ML) is additionally deposited without any sample annealing. 
This process induces defects in the middle of the indium wires (Fig. 2(b)). In the second conductivity measurement (Fig. 2(c), dotted line), $\sigma$ is significantly decreased compared to the first above 130 K, while the two measurements collapse to a single curve below 130 K. 
Because electron conduction is suppressed by defects in the wires, the difference between the two (Fig. 2(c), dotted-dashed line) is the contribution of the surface reconstruction. 
It gradually decreases with decreasing temperature, followed by a sudden drop at 130 K. 
Below 120 K, conduction of the indium wires ceases, showing that the system is in the insulating phase. 

Let us note that the electron transport phenomenon revealed here originates from the unique surface state of Si(111)-4$\times 1$-In reconstruction.
A thin indium film on a silicon surface with bulk-like electronic states would not exhibit this change.
The result is in agreement with a phase transition recently found by Yeom \textit{et al}.\cite{Yeom_In4x1} 
They clarified that the system undergoes a transition around 130 K accompanied by gap opening at the Fermi level, leading to periodic modulation of lattices and charges below the transition temperature.
Although its mechanism is still under debate, it is clear from our finding that the system undergoes a metal-insulator transition.

Although macroscopic surface electron conduction through steps and domain boundaries seems remarkable, this has been confirmed for the Si(111)-$\sqrt{3} \times \sqrt{3}$-Ag reconstruction in a similar configuration.\cite{Hasegawa_Agroot3}
The surface steps and domain boundaries may be the main sources of resistance in surface conduction, but the present experiment suggests that the conductivity is generally quite high as far as the surface reconstruction grows up to the steps.
However, disruption of the reconstruction near steps can significantly enhance their resistance.
This is possible if electron transfer between adjacent terraces is due to tunneling. In this case, only a small change in tunneling gap distance will lead to substantial enhancement of resistance.

\begin{figure}[tbp]
\begin{center}\leavevmode
\includegraphics[width=0.72\linewidth]{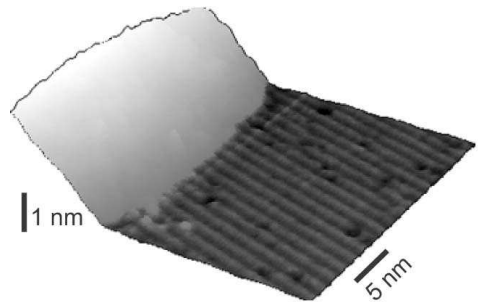}
\caption{
An STM image of the interface between the indium wires and a test tantalum pad.
The surface reconstruction has grown up to the edge of the pad. 
}
\label{fig3}
\end{center}
\end{figure}

One other factor significantly affects the global transport through the surface reconstruction: the electrode-indium wire interface. 
Figure 3 shows an STM image of the interface between the indium wires and a test tantalum pad. 
Surprisingly, the surface reconstruction has grown right up to the edge of the pad. This establishes the electrical connection between the electrodes and the wires. 
If silver rather than tantalum is deposited as electrodes on a sample after growth of the indium wires, wires near the electrode pads are destroyed, apparently by migrating silver atoms. 
We confirmed that this results in vanishing surface conduction. 
The growth of the indium surface reconstruction requires a clean silicon surface, which is maintained even near tantalum pads after flash cleaning.\cite{Dunn_TaPad}
This gives a natural explanation for the growth of the indium wires right up to the electrode interface. 

\begin{acknowledgments}
This work was partly supported by the Science and Technology Agency through Organized Research Combination System. 
The authors thank T. Yokoyama, L. D. Ueda-Sarson (NIMS), D. Pescia, A. Vaterlaus (ETH Zurich), and S. Hasegawa (University of Tokyo) for helpful discussions. 
\end{acknowledgments}

\newpage
a) Author to whom correspondence should be addressed. E-mail: UCHIHASHI.Takashi@nims.go.jp

b) Present address: Laboratorium f\"ur Festk\"orperphysik, Eidgen\"ossische Technische Hochschule (ETH) Z\"urich, CH-8093 Z\"urich, Switzerland.


\begin{references}
\bibitem{Lifshits_SurfacePhases}V. G. Lifshits, A. A. Saranin and A. V. Zotov, \textit{Surface Phases on Silicon} (Wiley, Chichester, 1994).

\bibitem{Henzler_SurfaceConduction}M. Henzler, in \textit{Surface Physics of Materials I}, edited by J. M. Blackely, (Academic Press, New York, 1975), pp. 241.

\bibitem{Henzler_AgSi}R. Schad, S. Heun, T. Heidenblut, and M. Henzler, Phys. Rev. B \textbf{45}, 11430 (1992).

\bibitem{Bauer_Pbfilm}M. Jalochowski, M. Hoffmann, and E. Bauer, Phys. Rev. B \textbf{51}, 7231 (1995).

\bibitem{Hasegawa_Agroot3}Y. Nakajima, S. Takeda, T. Nagao, and S. Hasegawa, and X. Tong,	Phys. Rev. B \textbf{56}, 6782 (1997).

\bibitem{Weitering_Si100}K. Yoo and H. H. Weitering, Phys. Rev. Lett. \textbf{87}, 026802 (2001).

\bibitem{Nogami_In4x1}J. Nogami, S. I. Park, and C. F. Quate,
	Phys. Rev. B \textbf{36}, 6221 (1987).

\bibitem{Abukawa_In4x1}T. Abukawa, M. Sasaki, F. Hisamatsu, T. Goto, T. Kinoshita, A. Kakizaki, and S. Kono,
 	Surf. Sci. \textbf{325}, 33 (1994).

\bibitem{Netzer_In4x1} J. Kraft, M. G. Ramsey, and F. P. Netzer, Phys. Rev. B \textbf{55}, 5384 (1997).

\bibitem{Yeom_In4x1}H. W. Yeom, S. Takeda, E. Rotenberg, I. Matsuda, K. Horikoshi, J. Schaefer, C. M. Lee, S. D. Kevan, T. Ohta, T. Nagao, and S. Hasegawa,
	Phys. Rev. Lett. \textbf{82}, 4898 (1999).

\bibitem{Bunk_In4x1}O. Bunk, G. Falkenberg, J. H. Zeysing, L. Lottermoser, R. L. Johnson, M. Nielsen, F. Berg-Rasmussen, J. Baker, and R. Feidenhans'l,
 Phys. Rev. B \textbf{59}, 12228 (1999).

\bibitem{Kumpf_In4x1}C. Kumpf, O. Bunk, J. H. Zeysing, Y. Su, M. Nielsen, R. L. Johnson, R. Feidenhans'l, and K. Bechgaard,
	Phys. Rev. Lett. \textbf{85}, 4916 (2000).

\bibitem{Another_Phase}Figure 1(a) contains another slightly thicker phase on the upper side of the step. This structure is different from the $\sqrt{7} \times \sqrt{3}$ reconstruction, and has been called 'striped structure'.\cite{Netzer_In4x1} In addition, more defects in indium wires are visible in Fig. 1(c) than in Fig. 1(a).

\bibitem{footnote1}Macroscopic electron conduction through the Si(111)-4$\times 1$-In is nearly isotropic despite its local 1D structure; domains with different wire directions coexist due to the three-fold symmetry of the reconstruction. The unit of surface conductivity is siemens (S), the same as that of conductance. This is because the conductance of a square sheet is independent of its size. 

\bibitem{Sze_Semiconductor}S. M. Sze, 	\textit{Physics of Semiconductor Devices, 2nd Edition} (John Wiley \& Sons, Inc., New York, 1981).

\bibitem{Handbook_ChemPhys}\textit{Handbook of Chemistry and Physics, 82nd Edition} (CRC Press, 2001).

\bibitem{Dunn_TaPad}A. W. Dunn, B. N. Cotier, A. Nogaret, P. Moriarty, P. H. Beton, and S. P. Beaumont,
	Appl. Phys. Lett. \textbf{71}, 2937 (1997).

\end{references}
\end{document}